\documentclass[twocolumn,tighten,times,pdflatex]{aastex631}
\usepackage{bm}
\usepackage{color}
\usepackage{graphicx}

\received{\today}

\submitjournal{ApJ}
\shorttitle{Rotational Dependence of Large-scale Convective Dynamo}
\shortauthors{Masada and Sano}
\begin{document}
\title{Rotational Dependence of Large-scale Dynamo in Strongly-stratified Convection: What Causes It ?}
\correspondingauthor{Youhei MASADA}
\email{ymasada@fukuoka-u.ac.jp}

\author[0000-0002-0786-7307]{Youhei Masada}
\affiliation{Department of Applied Physics, Fukuoka University, Fukuoka 814-0180, Japan}

\author[0000-0001-9106-3856]{Takayoshi Sano}
\affiliation{Institute of Laser Engineering, Osaka University, Suita, Osaka 565-0871, Japan}
\begin{abstract}
  In a rigidly-rotating magnetohydrodynamic (MHD) system with convective turbulences, a large-scale dynamo, categorized as the $\alpha^2$-type,
  can be excited when the spin rate is large enough. In this paper, the rotational dependence of the $\alpha^2$-type dynamo and the
  cause of it are explored by mean-field (MF) dynamo models coupled with direct numerical simulations (DNSs) of MHD convections. Bearing the
  application to the solar/stellar dynamo in mind, we adopt a strongly-stratified polytrope as a model of the convective atmosphere. Our DNS
  models show that the $\alpha^2$-type dynamo is excited when ${\rm Ro} \lesssim 0.1$ where ${\rm Ro}$ is the Rossby number defined with the
  volume-averaged mean convective velocity. From the corresponding MF models, we demonstrate that the rotational dependence of the $\alpha^2$-type
  dynamo is mainly due to the change in the magnitude of the turbulent magnetic diffusion. With increasing the spin rate, the turbulent magnetic
  diffusion weakens while the $\alpha$-effect remains essentially unchanged over the convection zone, providing the critical point for the excitation
  of the large-scale dynamo. The ${\rm Ro}$-dependence of the stellar magnetic activity observable in the cool star is also discussed from the view 
  point of the rotational dependence of the turbulent electro-motive force. Overall our results suggest that, to get a better grasp of the stellar
  dynamo activity and its ${\rm Ro}$-dependence, it should be quantified how the convection velocity changes with the stellar spin rate with taking
  account of the rotational quenching and the Lorentz force feedback from the magnetic field on the convective turbulence.
\end{abstract}

\section{Introduction}
Almost all of the late-type main-sequence (MS) stars, including our Sun, emit X-rays from a magnetically-confined plasma known as a corona.
Since the stellar coronae are thought to be heated by the release of magnetic energy generated by a magnetic dynamo process, the stellar X-ray
luminosity is considered as an important proxy for the stellar magnetic activity. A lesson of stellar observations is that the stellar X-ray
luminosity, i.e., the level of the magnetic activity, in solar and late-type stars is well-regulated by the spin period of the stars
\citep[e.g.,][]{pallavicini+81,noyes+84,pizzolato+03,wright+11}. Qualitatively, the younger stars with the shorter spin period, the more
magnetically-active it becomes. As the star ages and spins down due to the rotational braking \citep[e.g.,][]{skumanich72}, the stellar
magnetic activity becomes weaker. Actually, our Sun is an older star with a longer spin period compared to the other solar-type stars,
having a relatively lower level of the magnetic activity.

A key quantity to understand the physics lying behind the stellar rotation--activity relationship should be the ``Rossby number'',
${\rm Ro} \equiv P_{\rm rot}/\tau$, the ratio of the stellar rotation period, $P_{\rm rot}$, and the convective turnover time, $\tau$
\citep{noyes+84,pizzolato+03}. This is inferred by statistical studies on a large number of stellar samples with different masses,
spin periods and X-ray luminosities. For example, \citet{wright+11} shows, from a sample of $824$ solar and late-type stars, that the power-law
slope $L_X/L_{\rm bol} \propto {\rm Ro}^{\beta}$ is well-fitted by $\beta = -2.7$ in the range ${\rm Ro} \gtrsim {\rm Ro}_{\rm sat} \approx 0.1$,
while $\beta \simeq 0.0$, due to the saturation of $L_X$, for very fast rotators with ${\rm Ro} \lesssim {\rm Ro}_{\rm sat}$, where $L_X/L_{\rm bol}$
is the ratio of the stellar X-ray to bolometric luminosities. Note that the mass-dependent convective turnover time, based on the mixing-length
theory \citep[e.g.,][]{kippenhahn+90}, is adopted for the ${\rm Ro}$ in these observational studies. 

Not only for G-type and F-type dwarfs, composed of convective envelope and radiative core like the Sun, recent observations indicate that the
same rotation--activity relationship holds even for fully-convective M-type dwarfs \citep{wright+16,wright+18} and evolved stars in post-MS phases
\citep{lehtinen+20,godoy-rivera+21}. This implies that there exists a universal dynamo scaling and the underlying dynamo mechanism should be common
to late-type stars irrespective of their mass or evolutionary stage and the resulting different internal structures. However, theoretical dynamo 
models still do not have the ability to predict the rotational dependence of the magnetic activity quantitatively. 

After the first successful global simulation, presented by \citet{ghizaru+10}, that reproduced the solar-like cyclic large-scale magnetic field,
the numerical modeling of the solar and stellar dynamos has made significant progress in the past decade. See \citet{brun+17} and
\citet{charbonneau20} for reviews. While most of recent numerical models have examined the solar dynamo process in greater detail as a prototypical 
model for stellar dynamos \citep[e.g.,][]{kapyla+12,masada+13,nelson+13,fan+14,augustson+15,guerrero+16,hotta+16,kapyla+17}, few of them have focused
on the rotational dependence of the convective dynamo process. 

\citet{mabuchi+15} found that, in the regime ${\rm Ro} \lesssim 1$, the stellar differential rotation becomes weaker and the mean kinetic helicity
(net helicity) decreases with the ${\rm Ro}$, while the convective dynamo becomes more active as ${\rm Ro}$ decreases. \citet{strugarek+17} found
from a set of turbulent global simulations that the cycle period of the large-scale magnetic field scales as ${\rm Ro^{-1}}$, which is compatible
with the ``rotation--activity cycle'' relationship of the Sun and the other solar-type stars \citep[see also,][]{strugarek+18, warnecke+18}. With
focusing on the configuration of the stellar magnetic field, \citet{viviani+18} found numerically that an axisymmetric (non-axisymmetric) dynamo
mode dominates at slow (fast) rotation and the transition between them occurs at around ${\rm Co} \simeq 3.0$, where ${\rm Co} \propto {\rm Ro}^{-1}$
is the Coriolis number. From the viewpoint of the dynamo mechanism, \citet{warnecke+20} found that the trace of turbulent $\alpha$ tensor, which
induces an inductive effect, increases with ${\rm Co}^{0.5}$ for moderate rotation, and levels off for rapid rotation. 

The rotational effect on the convective dynamo has been studied in a Cartesian system as well. A pioneering work is \citet{kapyla+09} in which they
demonstrated, for the first time, that a rigidly-rotating convection can drive a large-scale dynamo even in the absence of the differential rotation.  
They also found that the turbulent diffusivity decreases monotonically with the spin rate while the turbulent $\alpha$-effect is almost independent
of the rotation. As a result, the large-scale dynamo is excited only when the system rotates rapidly enough. See, e.g., \citet{kapyla+13},
\citet{masada+14a,masada+14b} (hereafter, MS14a,b), and \citet{bushby+18} for the details of the large-scale dynamo in rigidly-rotating convections.

Although significant progress has been made in the theoretical study of the solar and stellar dynamos, we still do not have a solid explanation
on the physics lying behind the stellar rotation--activity relationship. To reveal it, we need to further understand how the rotation affects the
property of the convective dynamo. We study, in this paper, the rotational dependence of the large-scale dynamo in rigidly rotating convections.
Bearing the application to the solar/stellar dynamo in mind,  as with our previous work \citep[][hereafter MS16]{masada+16}, we adopt a
strongly-stratified polytrope as a model of the convective atmosphere, that is the main difference from the earlier studies in which the
weakly-stratified atmosphere is focused \citep[][MS14a,b]{kapyla+09}. A unique approach in our study is to utilize the mean-field (MF) dynamo
models coupled with direct numerical simulations (DNSs) of MHD convections to explore the cause of the rotational dependence of the large-scale
dynamo. See MS14b for the MF dynamo model linked to the corresponding DNS. 

The outline of this paper is as follows. In \S~2, a numerical setup and the model of the convective atmosphere are introduced. The results obtained
from our DNS runs are presented in \S~3. We construct, in \S~4, a MF dynamo model, using the DNS results as profiles of the turbulent electromotive
force (EMF). Then, we try to pin down the physics which dominates the onset and rotational dependence of the large-scale dynamo. After discussing
the stellar rotation--activity relationship from the viewpoint of the rotational dependence of the turbulent EMF, we summarize our findings in \S~5. 

\section{Numerical Setup}
Convective MHD dynamo system is solved numerically in Cartesian domain. Our model covers only the convection zone (CZ) of depth $d$ ($0 \le z \le d$),
where $x$- and $y$-axes are taken to be horizontal, and $z$-axis is pointing downward. We set the width of the domain to be $W = 4d$. See Figure~1(a)
for the setup of the simulation domain. 
\begin{figure}[ht!]
\epsscale{1.15}
\plotone{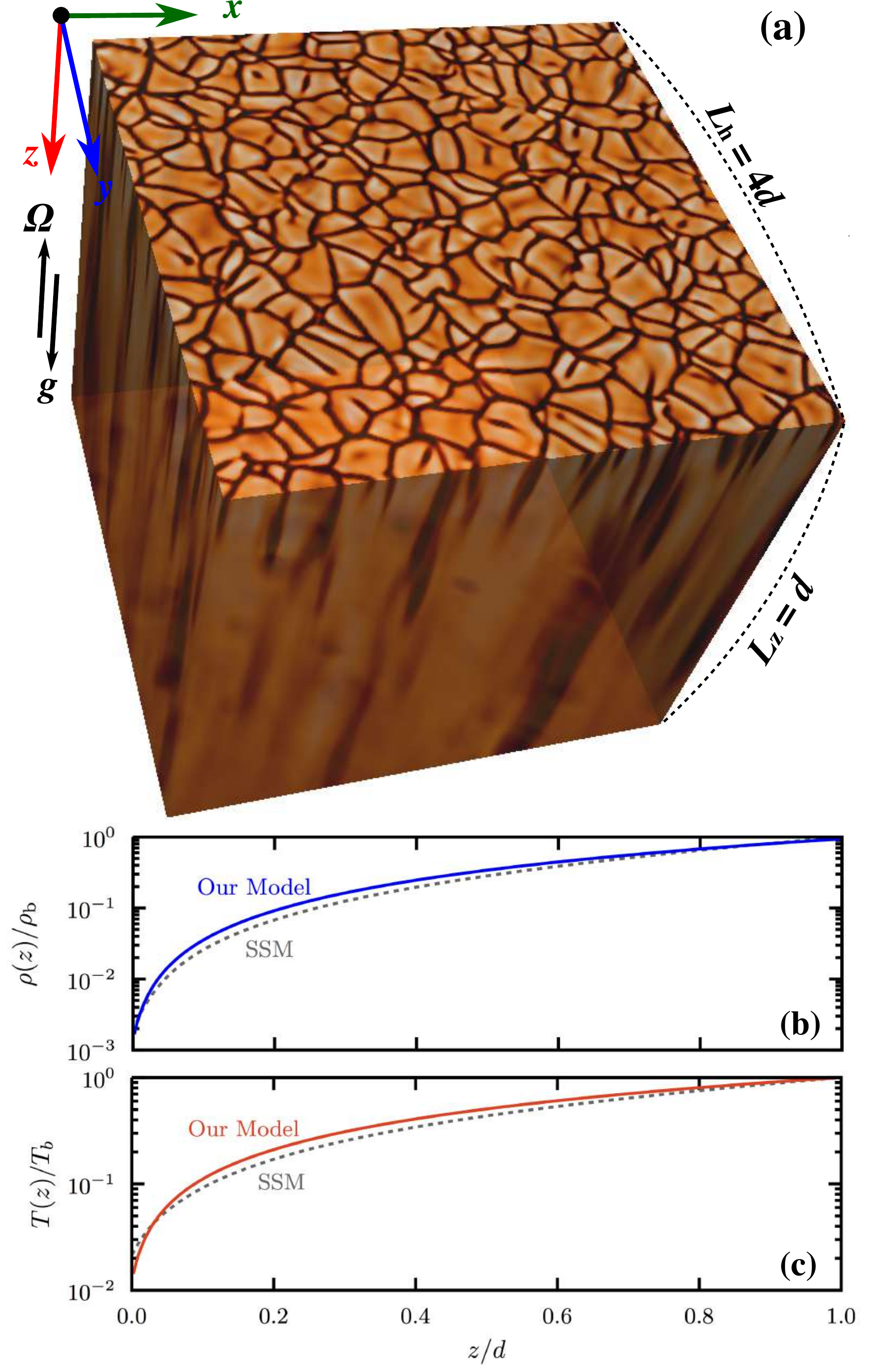}
\caption{(a) Setup of our simulation model. (b) and (c) are the initial profiles of the density (blue solid) and temperature (red solid).
  The dashed line in each panel is the profile of the standard solar model (SSM) in the range $0.71 \le r/R_\odot \le 0.991$ \citep[Model~S:][]{CD+96}.
  The profiles are normalized by the values at the bottom of the CZ. } 
\label{fig1}
\end{figure}
\begin{figure*}[htbp]
  \epsscale{1.15}
  \plotone{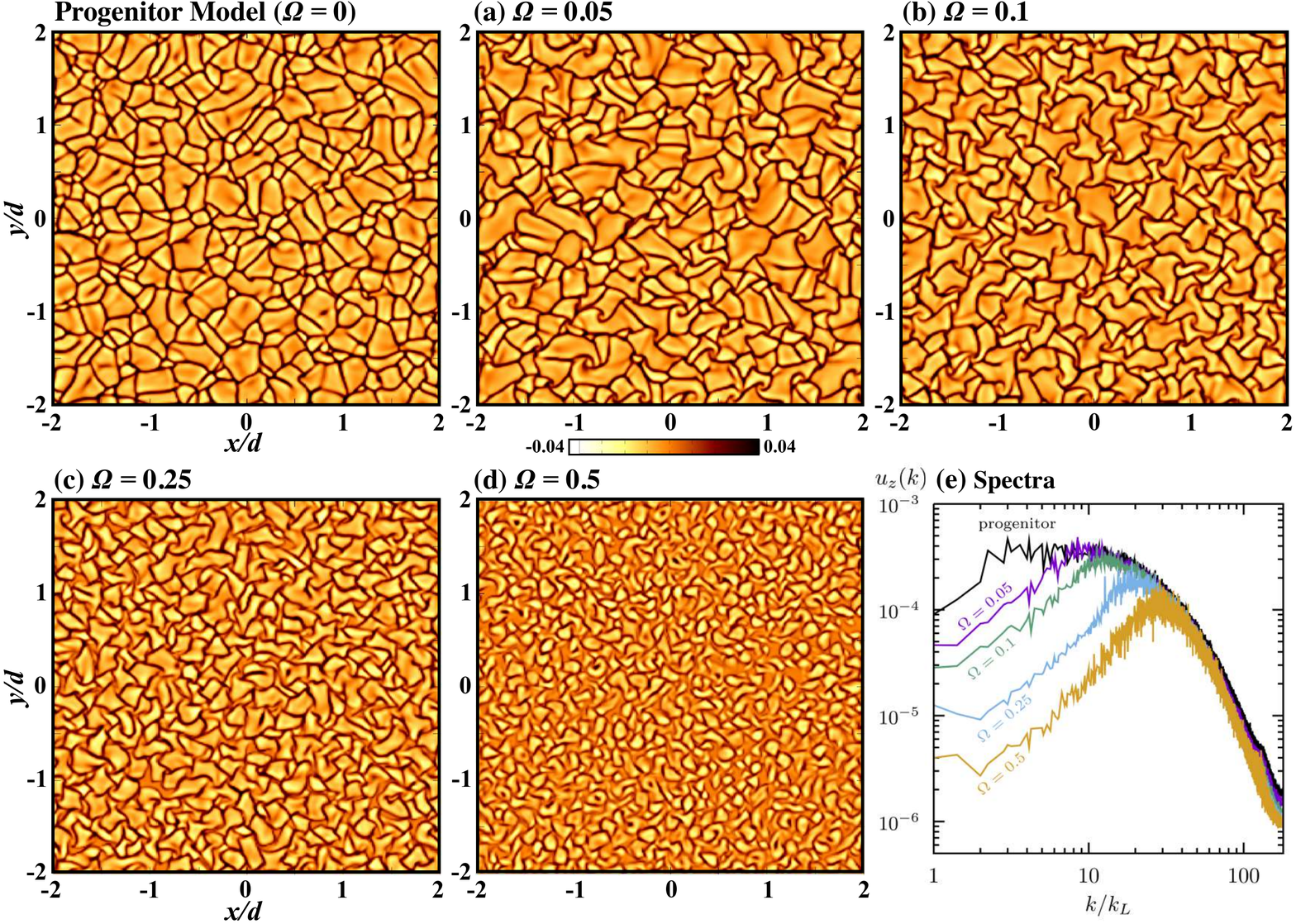}
  \caption{(a)--(f) Distributions of $u_z$ on the CZ surface at the saturated stage for the models with different $\Omega$. As for the comparison,
    $u_z$ for the progenitor model is shown in the top left panel. The saturated state is achieved after the relaxation phase of
    $t \lesssim 100\tau_{\rm cv}$, where $\tau_{\rm cv} = 31$, $34$, $44$, $54$ for the models~1--4. (e) Two-dimensional spectra of the vertical
    components of the convection velocity for the models with different $\Omega$.} 
  \label{fig2}
\end{figure*}

The following set of equations for fully-compressible MHD is solved in the rotating reference frame with the angular velocity of 
${\bm \Omega} = -\Omega {\bm e}_z$,
\begin{eqnarray}
\frac{\partial \rho}{\partial t} & = & - \nabla\cdot (\rho {\bm u})  \;, \label{eq1} \\ 
\frac{{\mathcal{D} }{\bm u}}{\mathcal{D} t}  & = & - \frac{\nabla P}{\rho}  
+ \frac{{\bm J}\times{\bm B} }{\rho} - 2{\bm \Omega}\times{\bm u} + \frac{\nabla \cdot {\bm \Pi}}{\rho} + {\bm g}  \;, \ \ \ \ \label{eq2} \\
\frac{\mathcal{D}\epsilon }{\mathcal{D} t} & = & - \frac{P\nabla\cdot {\bm u}}{\rho}  + \mathcal{Q}_{\rm heat} \;, \label{eq3} \\
\frac{\partial {\bm B} }{\partial t} & = & \nabla \times ({\bm u} \times {\bm B} - \eta_0 {\bm J})\;, \label{eq4}
\end{eqnarray}
where $\mathcal{D}/\mathcal{D}t$ is the total derivative, $\epsilon = c_{\rm V} T$ is the specific internal energy,
${\bm J} = \nabla \times {\bm B}/\mu_0$ is the current density with the vacuum permeability $\mu_0$, and ${\bm g} = g_0 {\bm e}_z$ is the gravity of
the constant $g_0$. The viscosity, magnetic diffusivity, and thermal conductivity are represented by $\nu_0$, $\eta_0$, and $\kappa_0$, respectively.
The magnitude of the angular velocity, $\Omega$, is the control parameter in this study. 

The viscous stress ${\bm \Pi}$ is written by ${\bm \Pi} = 2\rho \nu_0 {\bm S}$ with the strain rate tensor 
\begin{equation}
  S_{ij} = \frac{1}{2}\left( \frac{\partial u_i}{\partial x_j} + \frac{\partial u_j}{\partial x_i}
  - \frac{2}{3}\delta_{ij}\frac{\partial u_i}{\partial x_i} \right) \;,  \label{eq5} 
\end{equation}
The heating term $\mathcal{Q}_{\rm heat}$ consists of the thermal conduction, viscous and Joule heatings,
\begin{equation}
  \mathcal{Q}_{\rm heat} = \frac{\gamma\nabla\cdot(\kappa_0 \nabla\epsilon)}{\rho} + 2\nu_0 \bm{S}^2
  + \frac{\mu_0\eta_0\bm{J}^2}{\rho}\;, \label{eq6}
\end{equation}
We adopt a perfect gas law $P = (\gamma -1)\rho \epsilon$ with $\gamma = 5/3$ to close the equations.

We assume an initial hydrostatic equilibrium with a polytropic stratification given by 
\begin{equation}
\epsilon = \epsilon_0 + \frac{g_0z}{(\gamma-1)(m + 1)}\;, \ \ \ {\rm and} \ \ \ \rho = \rho_0 (\epsilon/\epsilon_0)^m \;,  \label{eq7}
\end{equation}
where $\epsilon_0$ and $\rho_0$ are the initial internal energy and density at the CZ surface. The polytropic index $m = 1.49$ is
adopted in our models,
providing the super-adiabaticity of $\delta\ \equiv \nabla - \nabla_{\rm ad}\ = 1.6 \times 10^{-3}$, where 
$\nabla_{\rm ad} = 1-1/\gamma$ and $\nabla = (\partial \ln T/\partial \ln P)$. Since the condition $\delta > 0$ is satisfied in whole the domain,
the convection in our system is essentially driven by the local entropy gradient, differing from the cooling-driven convection, qualitatively.
See, \citet{yokoi+22} for the difference between local and non-local (cooling) driven convections in turbulent transports
\citep[see also, e.g.,][]{spruit97,cossette+16}. 

Normalization quantities are defined by setting $g_0= \rho_0 = \mu_0 = c_p = d/2 = 1$. The normalized pressure scale-height at the surface,
defined by $\xi = H_p/d_{\rm cz} = (\gamma-1)\epsilon_0/(g_0d_{\rm cz})$, controls the stratification level and is chosen here as $\xi = 0.01$,
yielding a strong stratification with the density contrast between top and bottom CZs about $700$. 

Solid lines in Figures~1(b) and~1(c) show the profiles of the density and temperature for the initial setup of our simulation model. The density
and temperature profiles in the range $0.71 \le r/R_\odot \le 0.991$ of the standard solar model (SSM) are also shown by dashed lines in such a way
as to fit the computational domain. As the SSM, we adopt ``Model~S'' in \citet{CD+96}. Our model has a stratification almost encompassing the solar
convection zone except its uppermost part. 

All the physical variables are assumed to be periodic in horizontal directions. Stress-free boundary conditions are used in the vertical direction
for the velocity, perfect conductor and vertical field conditions are used for the magnetic field at the bottom and top 
boundaries, i.e.,
\begin{eqnarray}
  && \partial_z u_x = \partial_z u_y = u_z = 0 \ \ \ {\rm (top\ \& \ bottom\ BCs)} \;, \nonumber \\
  && B_x = B_y = \partial_z B_z = 0 \ \ \ \; \: {\rm (top\ BC)} \;, \nonumber \\
  && \partial_z B_x = \partial_z B_y = B_z = 0 \;  {\rm (bottom\ BC)} \;. \label{eq8}
\end{eqnarray}
A constant energy flux which prompts thermal convection is imposed on the bottom boundary, while the specific internal energy is fixed at the
top boundary,
\begin{eqnarray}
  \epsilon &=& \epsilon_0  \ \ \ \ \ \ \ \ \ \; \;  {\rm (top\ BC)} \;, \nonumber \\
  \partial_z \epsilon &=& {\rm const.} \; \ \; \: {\rm (bottom\ BC)} \;,  \label{eq9}
\end{eqnarray}

The fundamental equations are solved by the second-order Godunov-type finite-difference scheme that employs an approximate MHD Riemann solver
\citep{sano+98}. The magnetic field evolves with the Consistent MoC-CT method \citep{evans+88,clarke96}. Non-dimensional parameters ${\rm Pr} =20$,
${\rm Pm} =2$, ${\rm Ra} = 6\times10^7$, and the spatial resolution of ($N_x, N_y, N_z$) $=$ $(256,256,256)$ are adopted for all the models, where
the Prandtl, magnetic Prandtl, and Rayleigh numbers are defined by 
\begin{equation}
  {\rm Pr}  =  \frac{\nu_0}{(\kappa_0/\rho c_p )} \;,\ \ {\rm Pm} = \frac{\nu_0}{\eta_0}\;,
 \ \ {\rm Ra}  =  \frac{g_0 d^4}{\chi_0 \nu_0}\frac{\delta}{H_{p}}  \;, \label{eq10}
\end{equation}
where $\rho$, $\delta$ (super-adiabaticity), and $H_p$ are evaluated at the mid-part of the CZ, $z = d/2$. 

In the following, the volume-, and horizontal-averages are denoted by single angular brackets with subscripts ``\rm v", and ``\rm h", respectively.
The time-average of each spatial mean is denoted by additional angular brackets. The convective turn-over time and the equipartition field
strength are defined by $\tau_{\rm cv} \equiv 1/(u_{\rm cv}k_L)$ and $B_{\rm eq} (z) \equiv \sqrt{\langle \rho {\bm u}^2 \rangle_{\rm h}} $, 
where $u_{\rm cv} \equiv \sqrt{\langle \langle u_z^2 \rangle_{\rm v} \rangle}$ is the mean convective velocity at the saturated state and
$k_L \equiv 2\pi/d$ is the estimation on the scale of the energy-carrying eddies. Note that $B_{\rm eq}(z)$ is evaluated locally from the convective
kinetic energy, and thus has a depth-dependence. 

Since the sound speed in the deep CZ becomes very large in the strongly-stratified model and imposes a strict limit on computational time-step, a
long thermal relaxation time is required for our fully-compressible simulations. To alleviate it, we first construct a progenitor model, in which the
convection reaches a fully-developed state and the system becomes thermally-relaxed, by evolving a non-rotating hydrodynamic run for $800\tau_{\rm cv}$.
See, Figures 1b and 1c in MS16, as for our progenitor model. By imposing the rotation, which is the control parameter in this study, and then adding
a seed magnetic field to the progenitor model, the DNS run of the convective dynamo is started. The imposed angular velocities are $\Omega = 0.05$,
$0.1$~,$0.25$,~and $0.5$ for models~1,~2,~3, and~4, respectively. 

\section{Results of the DNS}
\subsection{Rotational dependence of convective motion}
\begin{figure*}[ht!]
  \epsscale{1.1}
  \plotone{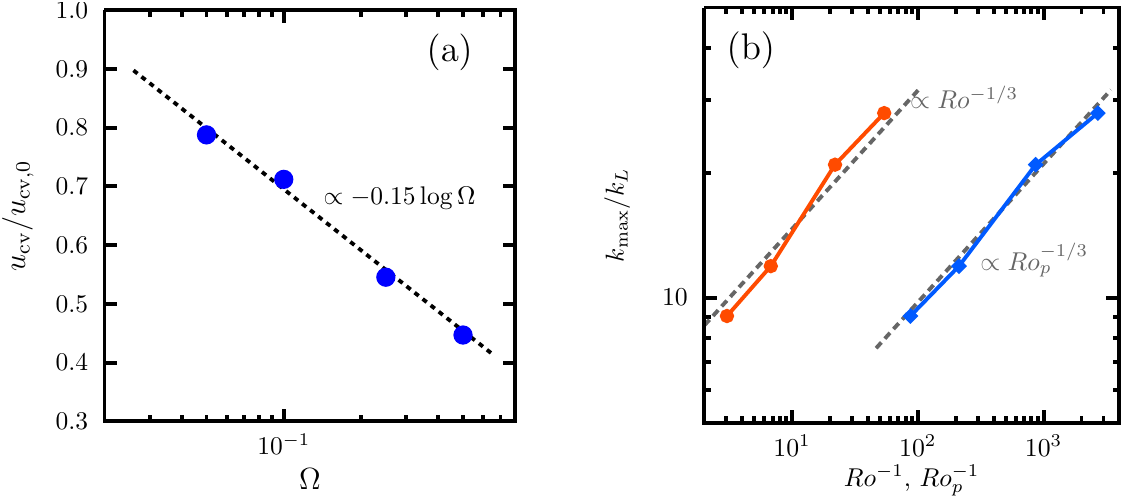}
  \caption{ (a)Dependence of normalized mean convection velocity $u_{\rm cv}/u_{\rm cv,0}$ on $\Omega$, where $u_{\rm cv,0} \simeq 0.013$ is the mean
    convective velocity of the progenitor model. (b)Dependence of $k_{\rm max}$, the wavenumber which gives the maximum of the spectra in Fig.2(e),
    on ${\rm Ro}$ (red) or ${\rm Ro_p}$ (blue). } 
  \label{fig3}
\end{figure*}
After the relaxation phase of $t \lesssim 100\tau_{\rm cv}$, each model reaches a saturated state. Note that the convective turn-over time is
different among models and is evaluated, at the saturated state, as $\tau_{\rm cv} = 31$, $34$, $44$, $54$ for models~1--4. See Fig.3(a) for the
typical convection velocity of each model. 

The distribution of $u_z$ on the horizontal
$x$--$y$ plane at the CZ surface is shown in Figure~2 for the saturated state of each model. The upper left panel demonstrates that for the
progenitor run (see, MS16). The panels~(a)--(d) are corresponding to the models~1--4 with $\Omega = 0.05,\ 0.1,\ 0.25$, and $0.5$, respectively. The
darker and lighter tones denote the downward and upward velocities.

The convective motion has a strong up-down asymmetry due to the strongly-stratified atmosphere: the broader and slower upflow-dominant cells
surrounded by narrower and faster downflow networks \citep[e.g.,][]{hurlburt+84,stein+98,miesch+00,brummell+02}. Since the rotation gives rise to
the Coriolis force acting on the convective motion, the size of the convective cell shrinks and thus the scale-separation becomes larger with the
increase of $\Omega$ \citep[see, e.g.,][]{brummell+02,miesch05}. 

The rotational quenching on the convective motion can be seen more quantitatively in the Fourier space. Shown in Figure~2(e) is the two-dimensional
spectra of the vertical component of the velocity for the models with different $\Omega$. The horizontal axis is  normalized by $k_L = 2\pi /d$.
Note that the spectrum of $\sqrt{u_z^2}$ at the each depth is projected onto a one-dimensional wavenumber $k^2 = k_x^2 + k_y^2$ and then is averaged
over the whole CZ in $z$-direction and given time span ($\simeq 20\tau_{\rm cv}$) at the saturated stage \citep[e.g.,][]{kapyla+09}. The purple, green,
cyan, and orange lines correspond to the models~1--4 with $\Omega = 0.05,\ 0.1,\ 0.25$, and $0.5$, respectively. The peak amplitude of the spectrum
becomes lower and the wavenumber which gives the maximum, $k_{\rm max}$, becomes larger with the increase of $\Omega$. 

As seen in the spectra, the higher the spin rate of the system, the lower the convection velocity becomes. Shown in Figure~3(a) is the dependence of
$u_{\rm cv}$ on $\Omega$, where $u_{\rm cv} \equiv \langle \langle u_z^2 \rangle_{\rm v} \rangle^{1/2}$ is the mean convective velocity. The vertical axis
normalized by that of the progenitor model, i.e., $u_{\rm cv,0} \simeq 0.013$. We can find that the convection velocity becomes lower with the increase
of $\Omega$, following a scaling law of $u_{\rm cv} \propto -\beta\log\Omega$ with $\beta = 0.15$. Since the compressible convection is accompanied by
contraction and expansion of fluid elements due to the change of the ambient density, the fluid motion is, even when the rotational axis and gravity
vector are aligned, more strongly constrained by the Coriolis force as $\Omega$ increases. This results in the lower convection velocity at the higher
$\Omega$ \citep[e.g.,][]{brummell+02,miesch05}. Actually, we have not figured out why $u_{\rm cv}$ decreases in proportion with $\log\Omega$, such a
dependence may be important to understand the why the ``saturation'' of the magnetic activity occurs in the rapidly-rotating regime of late-type
stars \citep[e.g.,][]{wright+11,wright+16}. We discuss the possible relationship between $\Omega$-dependence of $u_{\rm cv}$ and the saturation of the
stellar magnetic activity in \S~5.3. 

The effect of the rotation, i.e., the rotational constraint imposed by the Coriolis force, on the typical size of the convective cell is quantified.
Two kinds of the Rossby number are defined here in the following manner: 
\begin{equation}
  {\rm Ro} = \frac{u_{\rm cv}k_L}{2\Omega} \;,\ \ \ \ {\rm Ro}_p = \frac{u_{\rm peak}k_L}{2\Omega} \;, \label{eq11}
\end{equation}
where $k_L = 2\pi/d$, and $u_{\rm peak}$ is the convection velocity at the peak of each spectrum in Fig.2(e). Note that ${\rm Ro}$ can be evaluated as
$0.33, 0.15, 0.045, 0.019$ for the models~1--4. We put a special focus on this form of ${\rm Ro}$ in the following section. 

In Figure~3(b), the dependence of $k_{\rm max}$ on the inverse of the Rossby number is shown, where $k_{\rm max}$ is the wavenumber which gives the
maximum of the spectrum for each model in Fig.2(e). The red (blue) broken line with circles (diamonds) denotes the dependence on
${\rm Ro}$ (${\rm Ro_p}$). The gray dashed lines are reference power-law slopes with ${\rm Ro}^{-1/3}$ and ${\rm Ro}_p^{-1/3}$. 

$k_{\rm max}$ is found to change according to the scaling relation of $k_{\rm max} \propto {\rm Ro}^{-1/3}$ or
${\rm Ro}_p^{-1/3}$, which is consistent with the theoretically expected ``$-1/3$'' scaling for rotating hydrodynamic convection near onset. 
See \citet{chandra61} for the linear theory. While the dependence of $k_{\rm max}$ on ${\rm Ro}$ obtained in our DNS runs is compatible with that in
the other numerical models \citep[e.g.,][]{schmitz+10,stellmach+14,viviani+18,guervilly+19}, \citet{viviani+18} argue that the slope becomes steeper
($\propto Ro^{-1/2}$) in higher-resolution simulations at a rapid rotation regime \citep[see also,][]{featherstone+16}. This implies that the size of
convective cells might be more drastically reduced  if increasing the resolution and/or the spin rate even in our simulation models. 
\subsection{Rotational dependence of large-scale dynamo}
\begin{figure*}[ht!]
  \epsscale{1.0}
  \plotone{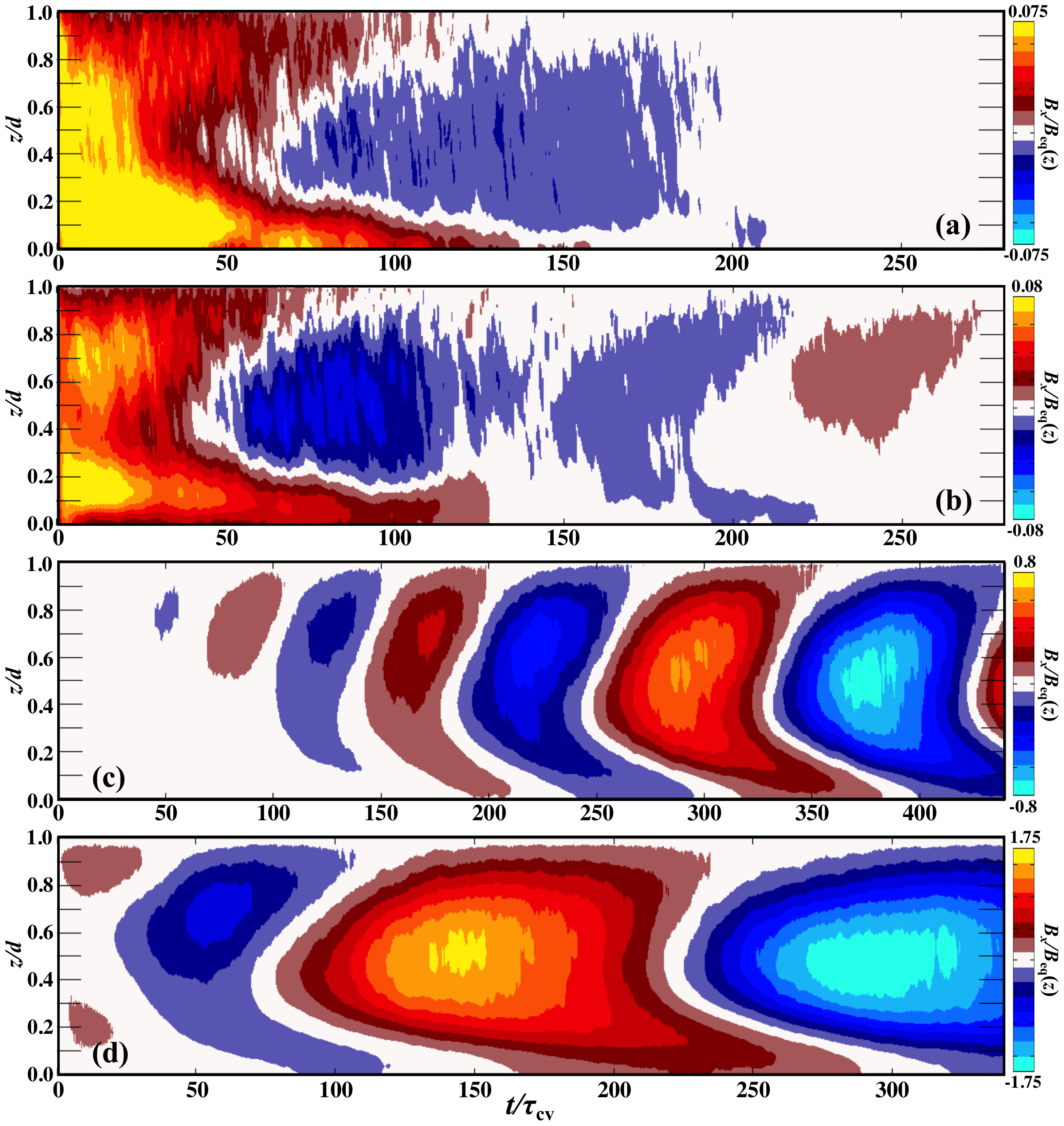}
  \caption{Time-depth diagram of normalized $\langle B_x \rangle_h$ for (a) model~1 ($\Omega = 0.05,\ {\rm Ro} = 0.33$),
    (b) model~2 ($\Omega = 0.1,\ {\rm Ro} = 0.15$), (c) model~3 ($\Omega = 0.25,\ {\rm Ro}=0.045$), and
    (d) model~4 ($\Omega = 0.5,\ {\rm Ro} = 0.019$). The red (blue) tone denotes the positive (negative)
    value of $\langle B_x \rangle_x$. The normalization unit for $\langle B_x \rangle_h$ is the equipartition field strength, $B_{\rm eq}(z)$.} 
  \label{fig4}
\end{figure*}
In MS16, we found that cyclic large-scale magnetic components are spontaneously organized
in a strongly stratified rotating convection. The model~4 with $\Omega =0.5$ in this study is corresponding one examined in MS16. When decreasing 
$\Omega$, we can see that the large-scale dynamo becomes gradually inactive and, is finally disappeared in the high ${\rm Ro}$ regime.

Figure~4 depicts the time-depth diagram of $\langle B_x \rangle_h$ for each model. The red (blue) tone denotes the positive (negative)
$\langle B_x \rangle_x$ in units of $B_{\rm eq}(z)$, where $B_{\rm eq}(z) = \langle \rho(z){\bm u}(z)^2 \rangle_h$ is the equipartition field strength.
The time is normalized by $\tau_{\rm cv}$ for each model. Since turbulent components of the magnetic field are averaged out by taking horizontal mean,
only the large-scale magnetic component and its time-space evolution can be observable in this figure. Note that $\langle B_y\rangle_h$ also shows a
similar cyclic behavior with $\langle B_x\rangle_h$, but with a phase delay of $\pi/2$, suggesting that an ``$\alpha^2$-type'' dynamo is excited in
our DNS models \citep[see, e.g.,][MS14a,b]{baryshnikova+87,radler+87,brandenburg+09}. See MS16 for the organization of the magnetic structure at the
CZ surface in the case of strongly-stratified atmosphere. 

In Figure~4, we can see two kinds of space-time evolution patterns: in the slowly rotating models with (a)$\Omega = 0.05$ and (b)$0.1$ (models~1~\&~2
with ${\rm Ro} = 0.33$ \& $=0.15$), the large-scale magnetic component starts to grow, but gradually stalls as time passes and finally disappear. On
the contrary to them, in the rapidly rotating models with (c)$\Omega = 0.25$ and (d)$0.5$ (models~3~\&~4 with ${\rm Ro} = 0.045$ \& $=0.019$), the 
oscillatory large-scale magnetic field is spontaneously organized. 

Commonly in the successful dynamo models, $\langle B_x \rangle_h$ has a peak in the mid-part of the CZ and migrates from there to top and bottom CZs.
The large-scale magnetic component becomes stronger with the increase of $\Omega$ (decrease of ${\rm Ro}$), reaching the super-equipartition in the
fastest spinning model. The cycle period of the large-scale magnetic component can be evaluated as $P_{\rm cyc} \simeq 100\tau_{\rm cv}$ for model~3 and
$\simeq 150\tau_{\rm cv}$ for model~4, implying that $P_{\rm rot}/P_{\rm cyc}$ is an increasing function of ${\rm Ro}$ (decreasing function of ${\rm Co}$). 
This seems to be compatible with the correlation between $P_{\rm rot}/P_{\rm cyc}$ and ${\rm Ro}$ found in the global model of the convective dynamo
\citep[c.f.,][and references therein]{warnecke18}. We will discuss the ${\rm Ro}$-dependence of the dynamo cycle in \S~5.2. 

Since the property of the large-scale dynamo changes between models~2 and ~3, the critical Rossby number that separates the success or failure of
the large-scale dynamo seems to exist, within the range $0.045 \lesssim {\rm Ro} \lesssim 0.15$, at around
\begin{equation}
  {\rm Ro}_{\rm crit} \simeq 0.1 \;. \label{eq12}
\end{equation}
Intriguingly, the ${\rm Ro}_{\rm crit}$, which is estimated from our DNS models in a Cartesian geometry, is compatible with that inferred from the
convective dynamo simulation in a rotating spherical-shell performed by, e.g., \citet{kapyla+12} and \citet{warnecke18} if we convert our Rossby 
number to their definition of the Coriolis number, despite the difference of the geometry. This implies that the presence of the large-scale flow,
such as the differential rotation and meridional flow, would not have a strong impact on the ``excitation'' of the large-scale dynamo. Specifically
in \S~5.1, we compare the ${\rm Ro_{crit}}$ (or ${\rm Co_{crit}}$) between our DNS models and the other global convective dynamo simulations.
\section{Analysis with Mean-field Dynamo Model}
\subsection{Governing Equation and Link to DNS}
\begin{figure}[htbp]
\epsscale{1.15}
\plotone{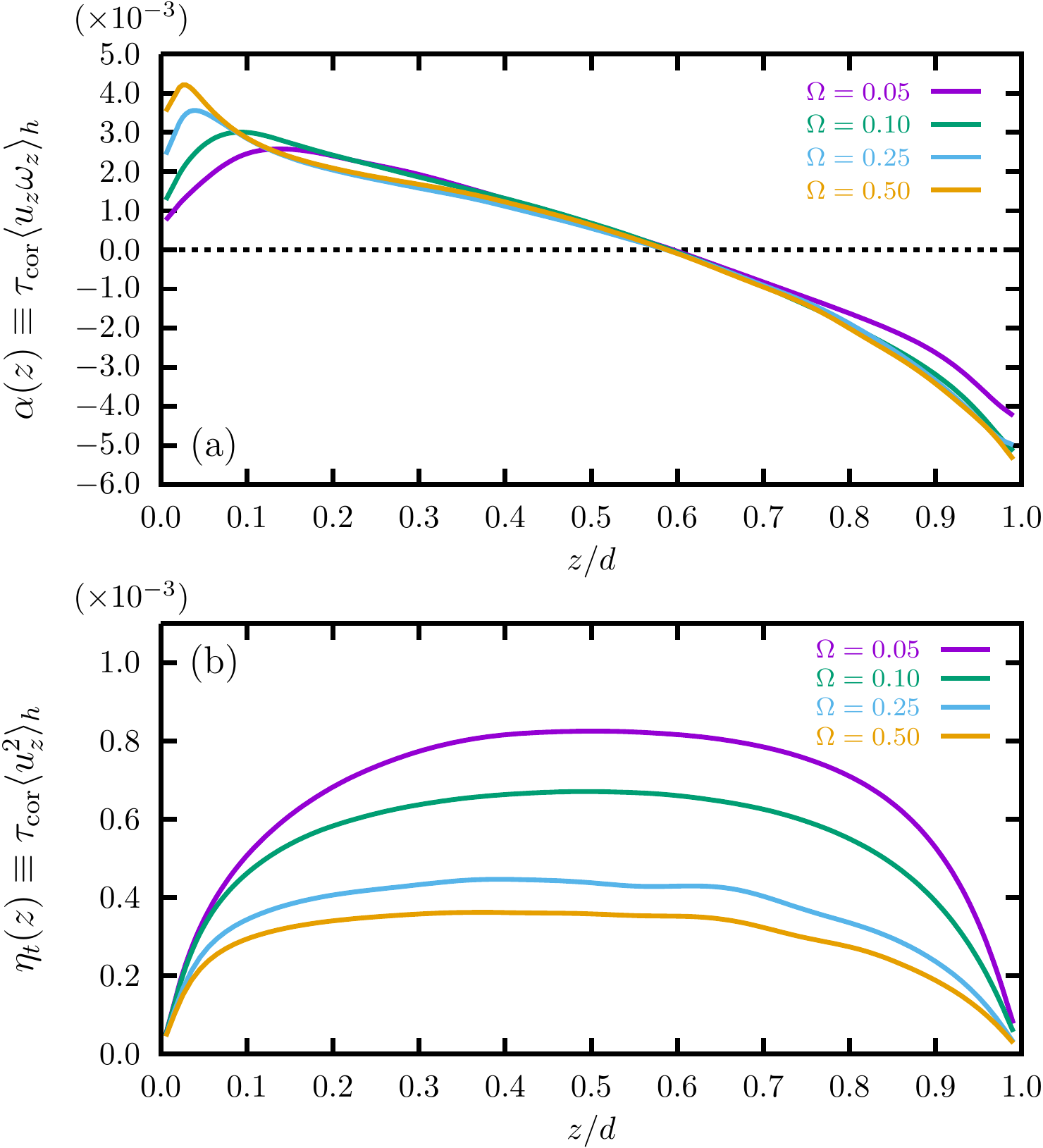}
\caption{Vertical profiles of (a)$\alpha_k(z)$ and (b)$\eta_t(z)$ extracted from the DNS results for each model. The different colors denote models
  with different $\Omega$.} 
\label{fig5}
\end{figure}
\begin{figure*}[htbp]
\epsscale{1.0}
\plotone{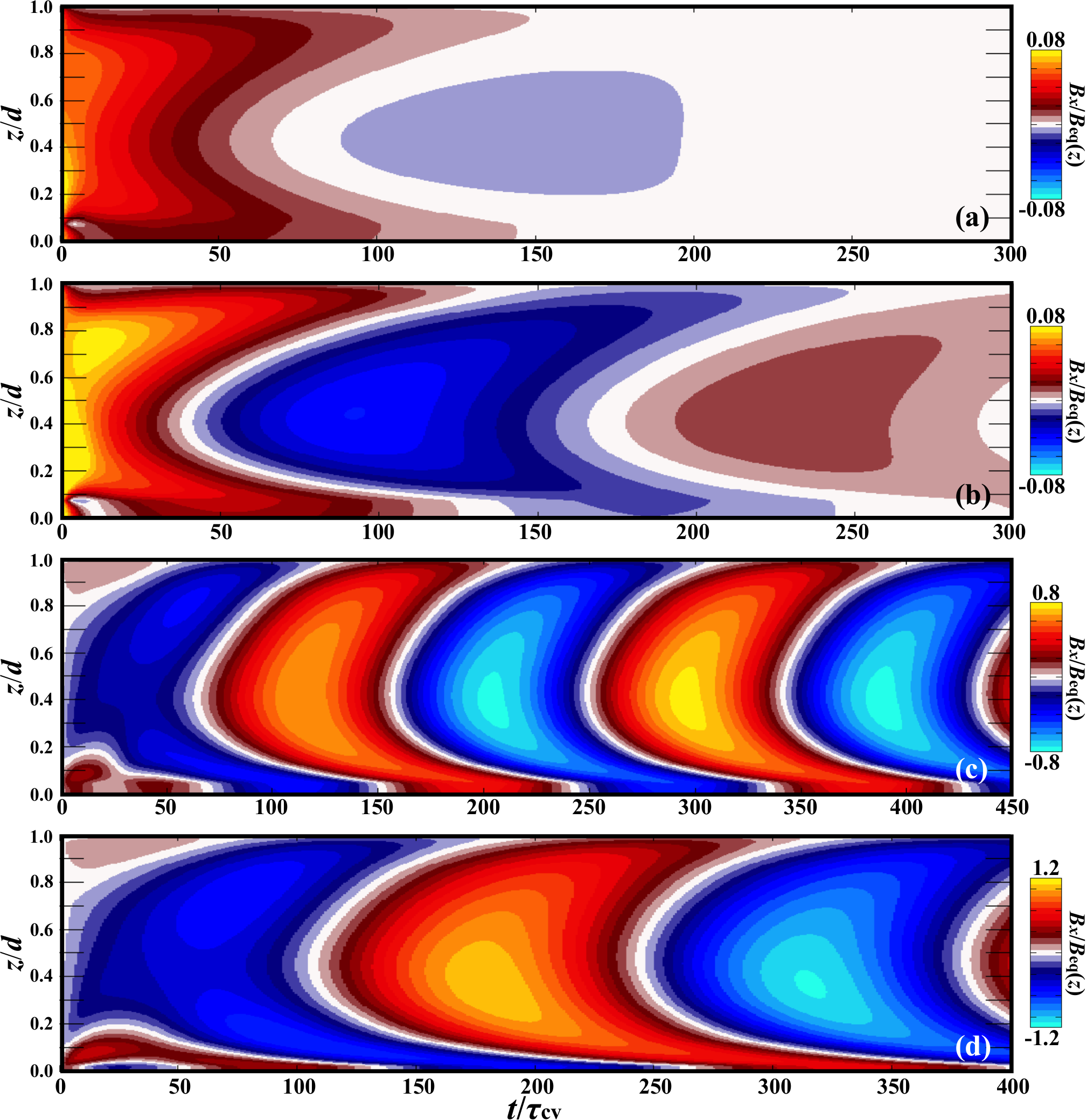}
\caption{Time-depth diagrams of $B_x^{\rm m}$ for the MF models corresponding to the DNS models with (a)$\Omega = 0.05$ (${\rm Ro} = 0.33$), 
  (b)$\Omega = 0.1$ (${\rm Ro} = 0.15$), (c)$\Omega = 0.25$ (${\rm Ro}=0.045$), and (d)$\Omega = 0.5$ (${\rm Ro} = 0.019$). The red (blue) tone
  denotes the positive (negative) value of $B_x^{\rm m}$. The normalization unit for $B_x^{\rm m}$ and $t$ are $B_{\rm eq}(z)$ and $\tau_{\rm cv}$
  evaluated from each DNS model.} 
\label{fig6}
\end{figure*}
In the previous section, we verified that the spin rate of the system is essential for exciting the large-scale dynamo. The critical Rossby number
that separates the success or failure of the large-scale dynamo, exists at ${\rm Ro}_{\rm crit} \simeq 0.1$. Then, a question naturally arises, 
``{\it How does the difference of the Rossby number impact on the physics of the convective dynamo and then change its property ?}''. 
To explore the answer to it, we utilize one-dimensional mean-field (MF) dynamo model, wherein the turbulent electromotive force (EMF) is determined
based on the velocity and helicity profiles directly extracted from the DNS models. 

Our MF dynamo model is constructed in a similar manner to \citet{masada+14b}. The $\alpha$-effect is solely responsible for the inductive effect in
our MF model, while the $\Omega$-effect is ignored there. This is because the angular momentum transport does not occur due to the alignment of the
rotation and gravity axes, and thus the differential rotation is not developed in our DNS setup. If the MF model linked to the DNS can reproduce the
rotational dependence of the large-scale dynamo observed in the DNS runs, it would be convincing evidence that the essential ingredient for it is
manifested in the turbulent EMF. Our aim is to confirm it. 

By dividing the variables into the horizontal mean and fluctuating components, $\bm{u} = {\bm{u}}_{\rm m} + \bm{u}'$ and
$\bm{B} = \bm{B}_{\rm m} + \bm{B}'$, the MF equation for the $\alpha^2$-type dynamo is obtained, from the induction equation, as 
\begin{equation}
  \frac{\partial \bm{B}_{\rm m}}{\partial t} = \nabla \times [ \bm{\mathcal{E}} - \eta_0 \nabla \times  \bm{B}_{\rm m} ] \;,
  \label{eq13}
\end{equation}
with
\begin{equation}
\bm{\mathcal{E}} =  \alpha \bm{B}_{\rm m}  - \eta_t \nabla \times \bm{B}_{\rm m} \;, \label{eq14}
\end{equation}
where $\eta_0$ is the microscopic magnetic diffusivity, $\bm{B}_{\rm m} = (B_x^{\rm m}, B_y^{\rm m})$ is the mean horizontal field, and
$\bm{\mathcal{E}}$ is the turbulent EMF \citep[e.g.,][]{parker79,krause+80,ossendrijver+02,brandenburg+05}. Note that $\bm{B}_{\rm m}$ has dependences
on time and depth. All the terms related to $\bm{u}_{\rm m}$ and $B_z^{\rm m}$ can be ignored in considering the symmetry of the system. The MF
coefficients $\alpha$ and $\eta_t$ represent the turbulent $\alpha$-effect and the turbulent magnetic diffusivity. The turbulent pumping effect is
ignored in this study for the simplicity. See, e.g., MS14b for the EMF with the turbulent pumping effect.  

The mean magnetic component generated by the large-scale dynamo produces a Lorentz force that tends to quench the turbulent motions, and thus control
the nonlinear saturation of the system. Since there is no definitive model to describe the quenching effect as yet, we adopt the prototypical model,
an algebraic $\alpha$-quenching, formulated as 
\begin{equation}
\alpha   =  \frac{\alpha_{k}}{1 + \langle \bm{B}_h\rangle_{\rm h}^2/B_{\rm eq}^2} \;, 
\end{equation}
\citep[see, e.g.,][for the review of the quenching models]{brandenburg+05}. 
Here, the subscript ``$k$" refers to the unquenched $\alpha$ coefficient, which is calculated directly from DNS results of the saturated
convective turbulence. The characteristic wavenumber
$k_c$ and the equipartition field strength at the saturated state, $B_{\rm eq,sat}$, are determined directly from DNS runs as 
\begin{equation}
k_c (z) = \frac{2\pi}{H_d(z)} \;, \ \ \ \ \ B_{\rm eq,st} (z) = \sqrt{\langle\langle \rho {\bm u}^2 \rangle_{\rm h}\rangle} \;, \label{eq16}
\end{equation}
where $H_d (z) = - {\rm d}z/{\rm d}\ln\langle\langle \rho \rangle_{\rm h}\rangle$ is the density scale height of the stratified atmosphere. 
With the first-order smoothing approximation (FOSA), the MF coefficients $\alpha_k$ and $\eta_{t}$ are determined, in anisotropic forms, by
\citep[e.g.,][]{kapyla+06a,kapyla+09}, 
\begin{eqnarray}
\alpha_k (z) & = &  - \tau_{\rm cor}[\langle\langle u_z\delta_x u_y \rangle_{\rm h}\rangle + \langle\langle u_x\delta_y u_z \rangle_{\rm h}\rangle] \equiv - \tau_{\rm cor}\mathcal{H}_{\rm eff} \;, \nonumber\\  
\eta_{t} (z)  & = &   \tau_{\rm cor}\langle\langle u_z^2 \rangle_{\rm h}\rangle \;,\ \label{eq17}  
\end{eqnarray}
where $\mathcal{H}_{\rm eff}$ is the effective helicity, and $\tau_{\rm cor}$ is the correlation time of the convective turbulence. 
With assuming
the Strouhal number is an order of unity in the CZ (${\rm St} = \tau_{\rm cor}\sqrt{\langle\langle u_z^2 \rangle_{\rm h}\rangle} k_c \approx 1$),
$\tau_{\rm cor}$ is determined, in the form with the depth dependence, by  
\begin{equation}
\tau_{\rm cor} = \tau_{\rm cor}(z) = \frac{1}{\sqrt{\langle\langle u_z^2 \rangle_{\rm h}\rangle} k_c} \;, \label {eq18}
\end{equation}
with $k_c$ defined in eq~(16).

The vertical profiles of (a)$\alpha_k(z)$ and (b)$\eta_t(z)$, which are determined from the DNS results of the saturated state, are
shown in Figure~5 for each model. The purple, green, cyan, and orange curves denote the models~1--4 with $\Omega = 0.05$, $0.1$, $0.25$, and $0.5$
(${\rm Ro} = 0.33$, $0.15$, $0.045$, and $0.019$), respectively. The time average spans in the range $220\tau_{\rm cv} \le t \le 250\tau_{\rm cv} $
for each model. Note that the profiles remain almost the same even if we change the range of time average.

It is intriguing that, with our definitions in eq~(17), the profiles of $\alpha_k$ are similar between models except the region around the CZ surface,
while the amplitude of $\eta_t$ decreases with the decrease of ${\rm Ro}$. $\alpha_{k}$ has a maximum at around the CZ surface and decreases deeper down
with the sign change at around the mid-part of the CZ. See, e.g., \citet{brummell+98} and \citet{miesch+00} for the depth-dependence of the kinetic
helicity induced by rotating convective turbulences. Qualitatively, the dependence of $\alpha_{k}$ or $\eta_t$ on the spin rate is compatible with that
seen in the model with the weakly-stratified convection \citep[][]{kapyla+09}. 

When seeing the dependence of effective kinetic helicity $\mathcal{H}_{\rm eff}$ on $\Omega$,
the larger the spin rate, the smaller the $\mathcal{H}_{\rm eff}$ becomes, throughout the CZ except near the surface\footnote{This tendency should be strongly
related to the formation of the convective column. In the regime of ${\rm Ro} \ll 1 $, the convective motion tends to align in columnar rolls
parallel to the rotation axis. In the convection columns, the velocity is mostly perpendicular to the rotation axis while the vorticity is mostly
parallel to the rotation axis, resulting in a small net kinetic helicity in the regime of the low ${\rm Ro}$ \citep[e.g.,][]{knobloch+81,mabuchi+15}.
It was also clearly demonstrated by \citet{sprague+06} with an asymptotic numerical analysis of unstratified turbulence that the largest net kinetic
helicity is established at a moderate rotation rate and decreases as rotation becomes even more rapid.}. On the other hand, the correlation time
becomes larger with the increase of $\Omega$ because $\langle\langle u_z^2 \rangle_h\rangle$ is a decreasing function of $\Omega$ (see Fig.3(a)).
As a result of cancellation of the $\Omega$-dependence between $\mathcal{H}_{\rm eff}$ and $\langle\langle u_z^2 \rangle_h\rangle$, the profile of $\alpha_k$
becomes almost independent of $\Omega$ in the most part of the CZ. 

At the top and bottom boundaries, we can anticipate that the convective turbulence is not fully developed, and thus $\alpha_k = \eta_t = 0$ should be
satisfied there. To smoothly vanish the turbulent EMF at the boundaries, we drop $\alpha_k$ and $\eta_t$ around there by extrapolating their profiles
by $\alpha_k'(z)=f(z)\alpha_k(z)$ and $\eta_t'(z) = f(z)\eta_t(z)$ artificially, with complementary error function, 
\begin{equation}
f(z) = \frac{1}{4} {\rm erfc}\left( \frac{z_b-z}{w_b}\right) {\rm erfc}\left( \frac{z-z_t}{w_t}\right) \;,\ \label{eq19}  
\end{equation}
where $z_i$ ($i=t, b$) represents the depth which separates the regions with and without fully developed turbulence, and $w_i$ ($i=t, b$) is the
width of the transition layer between them. We define $z_t$ and $z_b$ as the depth where $\alpha_k$ achieves the maximum and minimum values,
respectively. The width of the transition layer $w_i$ is an arbitrary parameter and determined here so that the amplitude of $\alpha_k$ drops fast
enough and becomes absolutely zero at top and bottom boundaries. 
By solving coupled
equations~(13)--(18) with substituting $\alpha_k'(z)$ and $\eta_t'(k)$ for $\alpha_k(z)$ and $\eta_t(k)$, we conduct a series of one-dimensional
simulation of the MF dynamo corresponding to each DNS model. The boundary conditions for the magnetic field in the MF dynamo model are same as those
adopted in the DNS runs (see, eq.(8)). 
\subsection{Results of MF dynamo model}
Shown in Figure~6 is the the time-depth diagram of $B_x^{\rm m}/B_{\rm eq}(z)$ for the MF model corresponding to each DNS model. Panels~(a)--(d) are
the MF counterparts of the DNS models~1--4 with $\Omega = 0.05$, $0.1$, $0.25$, and $0.5$ (${\rm Ro} = 0.33$, $0.15$, $0.045$, and $0.019$). The red
(blue) tone denotes the positive (negative) $B_x^{\rm m}$. Note that the normalization units $B_{\rm eq}(z)$ and $\tau_{\rm cv}$ are extracted from each
DNS model. Note that $B_y^{\rm m} $ also shows a similar cyclic behavior with $B_x^{\rm m}$, but with a phase delay of $\pi/2$ as in the DNS models. 

The property of the large-scale dynamo observed in the DNS models are reproduced, at least qualitatively, even in the simple MF dynamo model: in the 
MF counterparts to the slowly rotating DNS models, ${\bm B}_{\rm m}$ initially grows, but gradually decays as time passes, and finally disappear (see
panels~(a) and (b)). In contrast, in the MF models corresponding to the rapidly rotating DNS models, ${\bm B}_{\rm m}$ is maintained with a cyclic
periodicity (see panels~(c) and (d)). The oscillation period of ${\bm B}_{\rm m}$ in the MF model is also similar to that of
$\langle {\bm B}_h \rangle_{\rm h}$ in the DNS model. 

A remarkable difference between the MF and DNS models is the amplitude of the mean magnetic component in the fastest spinning model: while the field
strength reaches maximally $\sim 1.75B_{\rm eq}$ in the case of the DNS model at the mid part of the CZ, it shows $\sim 1.2B_{\rm eq}$ in the MF model.
On the other hand, when the saturation level is sub-equipartition (i.e., models~1--3), the amplitude of the large-scale magnetic component is almost
the same between MF and DNS models, regardless of whether the large-scale dynamo is maintained. This implies that some physics might be missing in the
quenching model, which controls the nonlinear saturation of the system, especially in realizing the regime of the super-equipartition field. 

Overall our results obtained by a series of one-dimensional simulation of the MF dynamo indicate that the space-time evolution of the large-scale
magnetic field in our setup can be understood as a manifestation of the propagation of ``dynamo wave'' due to the $\alpha^2$-type dynamo mode
\citep[e.g.,][]{baryshnikova+87,radler+87}. Furthermore, recalling the rotational dependence of the turbulent EMF, i.e., $\alpha_k$ and $\eta_t$ in
Fig.5, our results suggest that the ${\rm Ro}$-dependence of the large-scale dynamo in the rigidly-rotating convection is mainly due to the change
in the magnitude of the turbulent magnetic diffusivity: with increasing the spin rate, the turbulent magnetic diffusion becomes weaker while the
inductive $\alpha$-effect remains essentially unchanged over the CZ except near the surface, providing the ${\rm Ro}_{\rm crit}$ for the excitation of
the large-scale dynamo. 
\section{Discussion}
\subsection{Comparison of ${\rm Ro}_{\rm crit}$ with Global Simulations}
In \S~3.2, we found from the DNS results that there exists a critical Rossby number ${\rm Ro}_{\rm crit}$, which separates the success or failure of
the large-scale dynamo. Here we compare the ${\rm Ro_{crit}}$ with that obtained in the other convective dynamo simulations. As an example, we focus
on the global dynamo simulations performed with {\sc Pencil code} \citep[][]{pencil}. 

From the time-latitude diagram of the convective dynamo achieved in the global simulations with {\sc Pencil code}, we read that the critical Coriolis
number, ${\rm Co_{crit}}$, that separates the success or failure of the large-scale dynamo exits in the range
$4.7 \lesssim {\rm Co}_{\rm crit} \lesssim 7.6$ for \citet{kapyla+12} and
$4.5 \lesssim {\rm Co}_{\rm crit} \lesssim 6.6$ for \citet{warnecke18}, where the ${\rm Co}$ is defined there as, 
\begin{equation}
  {\rm Co} = \frac{2\Omega}{u_{\rm rms}k_f} \;. \label{eq20}
\end{equation}
with $u_{\rm rms} = \sqrt{(3/2)\langle \langle u_r^2 + u_\theta^2 \rangle_{\rm v}\rangle}$. When we convert our Rossby number, ${\rm Ro}$, to their
definition of the Coriolis number, with the relation $u_{\rm rms} \simeq \sqrt{3\langle \langle u_z^2 \rangle_{\rm v} \rangle} = \sqrt{3}u_{\rm cv} $,
our DNS models provides,  
\begin{equation}
3.8 \lesssim {\rm Co_{crit}} \lesssim 12.8 \;. \label{eq21}
\end{equation}

The condition~(21) for ${\rm Co_{\rm crit}}$ largely overlaps with that suggested from the results in \citet{kapyla+12} and \citet{warnecke18}, despite
the difference in the geometry of the simulation models. It is also interesting that, according to the result of \citet{kapyla+09}, even in the
large-scale dynamo operated in a weakly-stratified rigidly-rotating convection, the ${\rm Co_{\rm crit}}$ seems to be in the range
$4.1 \lesssim {\rm Co_{crit}} \lesssim 11.6$ (their density contrast of about $23$ across the domain). 

From the comparison between various models, it is suggested that the large-scale flow, such as the differential rotation and meridional flow, as well
as the stratificaton level, would not have a strong impact on the ``excitation'' of the large-scale dynamo, whereas the turbulent $\alpha$-effect
would be essential for it. Furthermore, this may imply that the ${\rm Ro_{crit}}$ (or ${\rm Co_{crit}}$) is universal in all the astrophysical dynamo
activites \citep[e.g.,][]{masada+22}. 
\subsection{{\rm Ro}-dependence of $P_{\rm cyc}/P_{\rm rot}$}
In \S~3.2, we roughly estimated the cycle period of the large-scale dynamo as  $P_{\rm cyc} \simeq 100\tau_{\rm cv}$ for model~3
and $\simeq 150\tau_{\rm cv}$ for model~4. Here we evaluate $P_{\rm rot}/P_{\rm cyc}$, where
\begin{equation}
\frac{P_{\rm rot}}{P_{\rm cyc}} = \left(\frac{2\pi}{\Omega}\right)\left(\frac{1}{P_{\rm cyc}}\right) \;, \label{eq21}
\end{equation}
and then compare it between our models and the other models of the global convective dynamo. 

Shown in Figure7 is ${\rm Ro}$-dependence of $P_{\rm rot}/P_{\rm cyc}$ for our DNS models. The red circle denotes the model with the successful
large-scale dynamo (models~3\&4), and the blue cross is the model with no dynamo solution (models~1\&2). For the reference, the low (high) ${\rm Ro}$
regime for the successful (unsuccessful) large-scale dynamo is red (blue) shaded. 

It seems that $P_{\rm rot}/P_{\rm cyc}$ for the successful dynamo model follows a scaling relation, $P_{\rm rot}/P_{\rm cyc}\propto {Ro}^\beta$ with
$\beta \simeq 1$. This interestingly suggests that, despite the difference in the geometry of the computational models, the scaling with respect to
${\rm Ro}$ shows a similar trend between our model and the global dynamo model \citep[][]{warnecke18}. However, our result is not statistically
significant because, in our simulations, in addition to the fact that only two models have periodic variations, both models have only been able to
calculate up to the early part of the cycle. To confirm the result, a larger sample of successful dynamo models and longer term calculations are
required. That is beyond the scope of this paper, and should be a target of our future work. 
\begin{figure}[ht!]
\epsscale{1.1}
\plotone{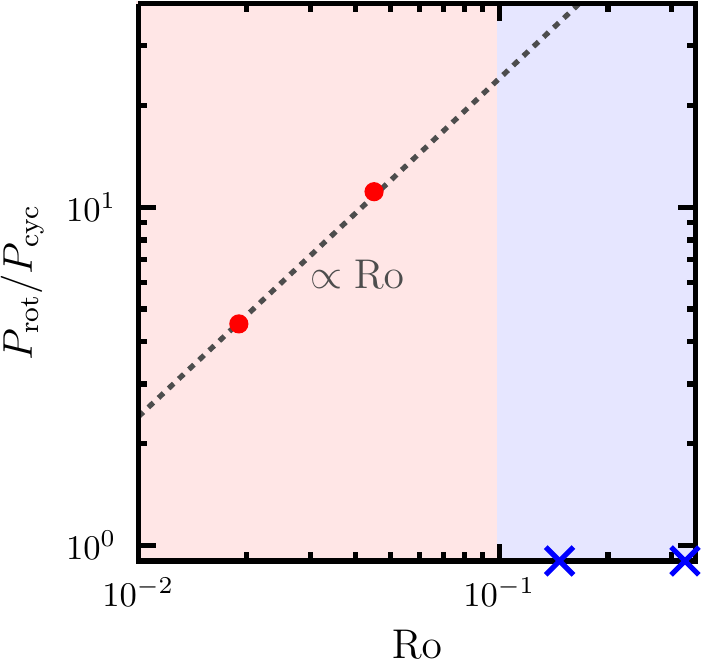}
\caption{Dependence of $P_{\rm rot}/P_{\rm cyc}$ on ${\rm Ro}$ for the models with $\Omega = 0.05$, $0.1$, $0.25$, and $0.5$ (${\rm Ro} = 0.33$, $0.15$,
  $0.045$, and $0.019$). The filled red circle presents $P_{\rm rot}/P_{\rm cyc}$ for the model with the successful large-scale dynamo (models~3\&4),
  while the blue cross is for the model with no dynamo solution (models~1\&2). The red shaded region denotes the low ${\rm Ro}$ regime
  (${\rm Ro} < {\rm Ro}_{\rm crit}$) for the successful large-scale dynamo, while the blue shaded region denotes the high ${\rm Ro}$ regime
  where the large-scale dynamo fails to be excited. } 
\label{f7}
\end{figure}
\subsection{A Clue to the Saturation of Stellar Magnetic Activity ?}
In \S~3.1, we found that the RMS value of the vertical component of the velocity, $u_{\rm cv}$,  becomes lower when increasing of $\Omega$ with a
scaling law of $u_{\rm cv} \propto -\log\Omega$. With leaving aside whether this functional form on $\Omega$ of $u_{\rm cv}$ is accurate, the response
of the convection velocity to $\Omega$ seen in Fig.3(a) may rather imply a sign that the convection velocity bottoms out in high $\Omega$ (low
${\rm Ro}$) regime. We will discuss here the possible relationship between the presence of the floor value of $u_{\rm cv}$ and the saturation of the
stellar magnetic activity via the dynamo number $N_{\rm D}$ in the framework of the MF dynamo model. 

With supposing the saturation of the convection velocity, we consider, as a demonstration, a simple piecewise function
\begin{equation}
\bar{u}_{\rm cv}(\bar{\Omega}) = \left\{ \begin{array}{l}
\bar{\Omega}^{-\delta} \ \ \ \ \ \ \ \ \ \ \ \ \ \ \; \; \; \; \; \;  ( \Omega \le \Omega_{\rm c} ) \\
\bar{u}_{\rm cv, b} (= {\rm const.})\ \  ( \Omega_{\rm c} < \Omega)
\end{array} \right.  \;, \label{eq23}
\end{equation}
where $\bar{u}_{\rm cv}$ and $\bar{\Omega}$ are the convection velocity and angular velocity normalized by arbitrary values, $\Omega_{\rm c}$ is an
arbitrary value of the angular velocity at which the dependence on $\Omega$ changes, $\bar{u}_{\rm cv,b}$ is an arbitrary floor value of the convection
velocity. As a measure of the buoyancy force with respect to the Coriolis force, we use the ``convective Rossby number'' defined by 
\begin{equation}
  {\rm Ro}_{\rm conv} = \bar{u}_{\rm MLT}/\bar{\Omega}
\end{equation}
where $\bar{u}_{\rm MLT}$ is the (normalized) typical convection velocity expected only from the stellar internal structure. Note that $u_{\rm MLT}$ is
usually evaluated based on the mixing-length theory (MLT) in which the impact of the rotational quenching on the convective motion is not taken
account of. 

When recalling the scalings $\eta_t \propto \tau_{\rm cor} u_{\rm cv}^2$ and $\tau_{\rm corr} \propto u_{\rm cv}^{-1}$ (see eqs.~(17) and (18)), the
normalized turbulent magnetic diffusivity $\bar{\eta}_t$ should be related to $\bar{u}_{\rm cv}$ as 
\begin{equation}
  \bar{\eta}_t(\bar{\Omega}) = \bar{u}_{\rm cv}(\bar{\Omega}) \;. 
\end{equation}
where the normalization unit for $\eta_t$ is arbitrarily chosen. When taking account of our insight that the inductive $\alpha$-effect is almost
independent of $\Omega$, the dynamo number for the $\alpha^2$-type dynamo mode is given
\begin{equation}
  N_{\rm D}(\bar{\Omega}) = \frac{\bar{\alpha}\bar{d}}{\bar{\eta}_t(\bar{\Omega})} \;,
\end{equation}
where $\bar{\alpha}$ is a constant $\alpha$-effect and $\bar{d}$ is the length scale normalized by an arbitrary value. We finally relate the 
luminosity due to the stellar magnetic activity ${L}_{\rm mag}$ to $N_{\rm D}$ via the functional form $L_{\rm mag} = N_{\rm D}^\chi$. 

Then, with eqs.(23)--(26), we show the dependence of $L_{\rm mag}$ on ${\rm Ro}_{\rm conv}$ as a function of $\chi$ in Figure~8. Here, as arbitrary values,
$\Omega_{\rm c} = \bar{u}_{\rm cv,b}=\bar{u}_{\rm MLT} = \bar{\alpha} = \bar{d} =1$ are chosen for the simplicity. As the power-law index $\delta$ in
eq~(23), we choose $\delta = 0.2$ which is evaluated from the fitting of the convection velocity for models~1 and~2 (unsuccessful dynamo models).
The different lines are corresponding to the cases with different values of $\chi$ ($\chi = 1$, $2$, $4$, $6$, $8$, and $10$).

As seen in this figure, with our toy model, the stellar magnetic activity possibly shows a power-law dependence
$L_{\rm mag} = {\rm Ro}_{\rm conv}^{-\delta\chi}$ in the high ${\rm Ro}_{\rm conv}$ regime, while it saturates in the low ${\rm Ro}_{\rm conv}$ regime. This 
implies that there exists a possible relationship between the presence of the floor value of $\bar{u}_{\rm cv}$ and the saturation of the stellar
magnetic activity. On the other hand, our model additionally suggests that, to explain the observed power-law slope $ \sim {\rm Ro}^{-2.7}$
\citep[e.g.,][]{wright+11,wright+16}, $L_{\rm mag}$ should have an extremely strong dependence on $N_{\rm D}$, i.e., $\chi \gtrsim 10$.

Although $N_{\rm D}$ determines the kinematic cycle period and growth rate, it may not determine the nonlinear cycle period nor the saturated magnetic
field strength \citep[e.g.,][]{tobias98,blackman+15}. It should be an important issue to understand how the strength of the dynamo-generated field
and the level of the magnetic activity depend on $N_{\rm D}$. To validate the effectiveness of our toy model and to get a better grasp of the stellar
dynamo activity and its ${\rm Ro}$-dependence, it should be quantified how the convection velocity changes with the stellar spin rate with taking
account of the rotational quenching and the Lorentz force feedback from both the small-scale \citep[e.g.,][]{hotta+21} and large-scale magnetic fields
on the convective turbulence.
\begin{figure}[ht!]
\epsscale{1.1}
\plotone{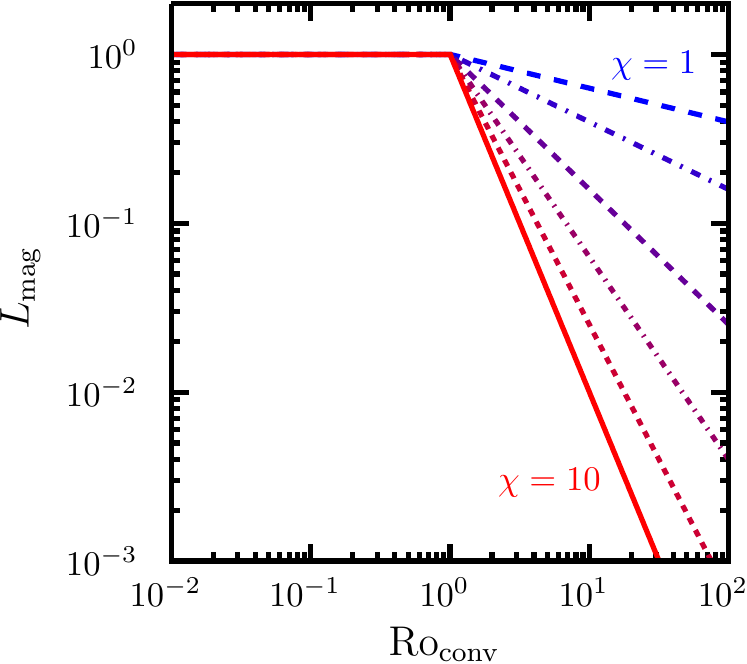}
\caption{Dependence of $L_{\rm mag}$ on ${\rm Ro_{\rm conv}}$ for the toy model of $L_{\rm mag} = N_{\rm D}^\chi$ with $\chi = 1$, $2$, $4$, $6$, $8$ and
  $10$. As arbitrary values, $\Omega_{\rm c} = \bar{u}_{\rm cv,b}=\bar{u}_{\rm MLT} = \bar{\alpha} = \bar{d} =1$ are chosen for the simplicity. As the
  power-law index $\delta$ in eq~(23), we choose $\delta = 0.2$ which is evaluated from the fitting of the convection velocity for models~1 and~2
  (unsuccessful dynamo models). } 
\label{f8}
\end{figure}
\section{Summary}
In this paper, we studied the rotational dependence of the large-scale dynamo in the rigidly rotating convection. Aiming at the application to the
solar and stellar dynamos, a strongly-stratified polytrope is adopted as the model of the convective atmosphere. Additionally to the DNS models, we 
took a unique approach to utilizing the MF model linked to the DNS to pin down the cause of the rotational dependence of the large-scale dynamo.
Our main findings are summarized as follows:\\

1. Due to the rotational constraint on the convective motion, the convection velocity decreases with the increase of $\Omega$. In addition, the
typical size of the convective cell $l$ is found to change according to the scaling relation of $l \propto {\rm Ro}^{1/3}$, consistent with the
theoretically expected scaling for rotating hydrodynamic convection near onset.\\

2. There exist two kinds of time-space evolution patterns in the large-scale magnetic field in our DNS models: in the slowly rotating models, the
large-scale magnetic component starts to grow, but gradually decays as time passes and finally disappear. In contrast, in the rapidly rotating models,
the oscillatory large-scale magnetic component is spontaneously organized. It becomes stronger with the decrease of ${\rm Ro}$, reaching the
super-equipartition in the fastest rotating model. The Rossby number for the successful large-scale dynamo is 
${\rm Ro} \lesssim {\rm Ro}_{\rm crit} \approx 0.1$. From the rough estimation on the dynamo cycle, a scaling relation 
$P_{\rm rot}/P_{\rm cyc} \propto {\rm Ro}$ can be found.  \\

3. The depth dependence of turbulent $\alpha$-effect, extracted directly from each DNS model, is similar despite the difference of ${\rm Ro}$ except
the region around the CZ surface. It peaks at around the CZ surface, decreasing deeper down with the sign change at around the mid-part of the CZ. In
contrast, while the vertical profile of the turbulent magnetic diffusion is similar between models, the amplitude is suppressed with the
decrease of ${\rm Ro}$. \\

4. Overall properties of the large-scale dynamo seen in the DNS models were reproduced in our simplified MF models at least qualitatively. 
This suggests that the ${\rm Ro}$-dependence of the convective dynamo is mainly due to the change in the magnitude of the turbulent magnetic diffusion:
with decreasing ${\rm Ro}$, the turbulent magnetic diffusion weakens while the inductive $\alpha$-effect remains essentially unchanged over the CZ, 
providing ${\rm Ro}_{\rm crit}$ for the excitation of the large-scale dynamo and the ${\rm Ro}$-dependence of dynamo activities. \\

Our finding, which brings up the possibility that the $\alpha$-effect is independent of the stellar spin rate, is suggestive and is widely applicable.
For instance, when modeling the stellar dynamo activity with a sort of mean-field model, a major issue in the conventional approach is how to give the
amplitude and profile of the $\alpha$-effect. However, if the $\alpha$-effect is independent of the stellar spin rate, there may be no need to worry
about it. Once the profile of $\alpha$-effect as shown in Fig.5(a) is given in the polar region and its latitudinal dependence is complemented by a
certain function (probably $\cos\theta$, with co-latitude $\theta$)\citep[e.g.,][]{kapyla+06b}, it may give a universal function of the $\alpha$-effect
applicable to all the G-type stars including the Sun. The next target of our research is to construct the global MF model for the solar and stellar
dynamos with $\alpha$-effect and turbulent magnetic diffusivity extracted from our DNS models studied here. 
\begin{acknowledgments}
  We thank J. Matsumoto, T. Takiwaki and N. Yokoi for fruitful discussions. This work has been supported by MEXT/JSPS KAKENHI grant Nos.
  JP18H01212, JP18K03700, JP21K03612. This work was partly performed under the joint research project of the Institute of Laser Engineering,
  Osaka University. Numerical computations were carried out on Cray XC50 at Center for Computational Astrophysics, National Astronomical
  Observatory of Japan.
\end{acknowledgments}

\end{document}